# DETECTION OF WORDPRESS CONTENT INJECTION VULNERABILITY

Md. Maruf Hassan[1], *Kaushik Sarker[2], Saikat Biswas[3], Md. Hasan Sharif[4]

Cyber Security Centre, Department of Software Engineering, Daffodil International University, Dhaka-1207, Bangladesh

*ABSTRACT*

*The popularity of content management software (CMS) is growing vastly to the web developers and the business people because of its capacity for easy accessibility, manageability and usability of the distributed website contents. As per the statistics of Built with, 32% of the web applications are developed with WordPress(WP) among all other CMSs [1]. It is obvious that quite a good number of web applications were built with WP in version 4.7.0 and 4.7.1. A recent research reveals that content injection vulnerability was found available in the above two versions of WP [2]. Unauthorized content injection by an intruder in a CMS managed application is one of the serious problems for the business as well as for the web owner. Therefore, detection of the vulnerability becomes a critical issue for this time. In this paper, we have discussed about the root cause of WP content injection of the above versions and have also proposed a detection model for the given vulnerability. A tool, SAISAN has been implemented as per our anticipated model and conducted an examination on 176 WP developed web applications using SAISAN. We achieved the accuracy of 92% of the result of SAISAN as compared to manual black box testing outcome.*

## 1. INTRODUCTION

In modern time, more than 3.6 billion people all over the world are using internet through web applications via a variety of different devices [3]. At present, to automate the existing manual processes of the basic activities of our daily life, web application is considered as the first step. Thus, all organizations in each sector have the propensity to restructure their working processes through web application for better performance and easy accessibility of their target users.

Web application become a compulsory needs for the businesses as well as for the people of this current era for its enourmous usefulness. However, the risk of exploitation has also be increased at the same time due to the existance of vulnerabilities in those web applications for insecure design and careless coding by the developers. According to OWASP and SANS, the most common vulnerability of the web applications are SQLi, Broken Authentication and Session Management, XSS, CSRF, Security Misconfiguration, LFI, LFD, Unprotected APIs, Buffer Overflow, etc. [4][5].

In recent years, much efforts have been initiated to identify those problems and have proposed different solutions to manage those problems. For instance, organizations such as MITRE, SANS and OWASP have developed and conducted security awareness programs to help organizations to mitigate those risks. However, despite these efforts, a recent study [6] shows that application developers still are unable to implement the effective countermeasures for web application's vulnerabilities. The frequent attack on web applications has been faced by the application owner because of the weakness present in web application and the reasons behind on most cases are the lack of knowledge in secure designing and coding practices.

Recently, CMS has become very popular for developing web applications. Being a reliable and well-known CMS the use of WordPress is increasing widely all over the world. A recent research has found vulnerability called WordPress content injection vulnerability in the most popular WordPress CMS version 4.7.0 and 4.7.1. However this vulnerability has already caused harm in web based informative sectors. From some previous study through internet, it is found that more than 20 black hat hacking groups are randomly defacing all kind of web applications where this vulnerability exists for a short period of time. A general statistics from the web source is given below on those hacking performed [7].

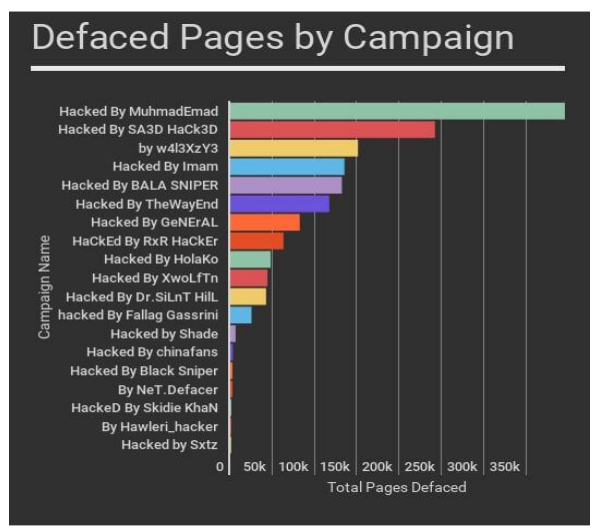

Figure 1. Statistis on wordpress content injection vulnerability defaced by campaign [7]

In order to find out those vulnerability in an automated way, different researches have been proposed verity of models and some of them have developed scanning or detection tools based on their proposed models. We also have performed some comparison study to compare the effectiveness between Manual and our developed tools.

In our research we have developd a detection tool that will help to detect such vulnerability. In this paper we have discussed briefly about WordPress content injection vulnerability and our proposed detection model along with the prevention techniques. We have prepared a tool, SAISAN using the given model and conducted an study on 176 WP developed web applications to validate the accuracy.

This paper is organized in ten sections. After the Introduction we have discussed the Background and related work in second section. Section three tells about our proposed model whereas the next section discusses about the WordPress content injection process. In section five, effect of WordPress content injection vulnerability has been discussed. Section six, seven and eight represent our Methodology, Result analysis and Discussion respectively. Finally, we have presented the Prevention techniques in section nine and we have concluded in section ten.

## 2. BACKGROUND AND RELATED WORKS

The invention of CMS creates the new opportunity for the developers in producing and managing digital contents in the web applications. As per statistics [8] more than 21 million web applications have been built using CMS like WordPress, Jumla, and Drupal. According to web technology survey, WordPress has become the most popular CMS at this era [9]. R. Connell.,(2013) performed another survey that reveals 54% of CMS user are satisfied over non-CMS user due to its cost effectiveness and flexibility [10].

Couple of studies has been conducted on different types of WordPress vulnerabilities as well as the weakness of its plug-ins. S. Patel et al.,(2013) conducted different web application attack e.g. sql injection, XSS, CSRF, LFI, etc. on three CMS versions i.e. joomla 1.6.2, drupal 6.22 and WordPress 3.2.1 built applications. Researchers found that direct security breaches of those applications were difficult but it was exploitable in most cases because of using third party vulnerable plug-ins where the possible sensitive directories information had been disclosed [10]. T. Koskinen et al.,(2013) ran an study on 322 Word Press plug-ins and discovered 860 vulnerabilities in 127 plug-ins where 72% were found as XSS vulnerability. This investigation did not find any clear connection between user rating plug-ins and vulnerable plug-ins [11]. R. Kumar.,(2016) provided and implemented 6 step methodology for developing WordPress website. After implementation of that application, a thorough security audit was performed where some vulnerabilities were found i.e. improper session management, stored XSS, file upload vulnerability, HTML injection etc. In this paper, the author provided a security guideline to fix those audit findings [12].H. Lim et al., (2016), performed an analysis on most popular CMSs like, WordPresss, joomla, Drupal, XE etc. They found that WordPress always holds number one position for popularity as well as vulnerability occurrence and proposed a guideline for content security which can be placed into CMS [13].An examination was performed by S. Lemeš et al.,(2011) on Joomla CMS vulnerability with default setting and shown that ISO/IEC 27000 series of standards can be used to improve security of Joomla based web portals [14].U. Ituen and M. Mukeshkrishnan.,(2015) formulated the most common and dangerous web application vulnerabilities and also exploitation procedure against those vulnerabilities. By analyzing the source code of the scripts it shows the way to breach confidential data from a web application by exploiting those vulnerability [15]. S. Stamm et al.,(2010) developed a add-ons based on content security policy (CSP) which works for alerting XSS and CSRF vulnerability of a web application. The add-ons was interacting with the browser in lower rate [16].

H. Trunde and E. Weippl.,(2015) figured out that automated black box testing is the best for crawling but it preforms poorly to detect new vulnerability whereas manual testing provides better output for all types of vulnerability detection [17]. For efficient vulnerability detection, various model and tools have been proposed and developed by the researchers. G. Agosta.,(2012) developed a tool for SQL injection (SQLi) and cross site scripting vulnerability detection based on symbolic code execution for PHP web applications. Result of the tool was compared with open source and commercial detection tools using SAMATE NIST benchmark. In their comparison, they also consider the false positive and false negative result analysis among all outputs [18]. N. Antunes et al.,(2009) proposed an approach to compare the effectiveness of vulnerability detection tools. That approach was used to define a concrete benchmark for SQL Injection vulnerability detection tools [19]. J. Bau.,(2010) carried out a study on the effectiveness of eight renowned web vulnerability scanners for some specific vulnerabilities. They found that the given tools were not performed well in detecting stored XSS and SQL injection. They also claimed that those tools would not even be able to detect the existence of the malware [20]. B. Delamore and R. Ko.,(2014) proposed a tool, Escrow by which it can detect SQL Injection for a large-scale web application together in efficient manner. This proposed tool was light-weight and platform-independent. Escrow uses a custom search implementation together with a static code analysis module to find potential target web applications [21]. Z. Djuric.,(2013) developed SQLIVDT for detecting SQLi weakness and performed vulnerability test using three vulnerable web applications. The result was compared with six well known vulnerability scanning tools for proving the SQLIVDT's accuracy [22]. N. Daud et al., (2014), executed a vulnerability analysis on an organization's information system using Nessus, Acunetix Web Vulnerability Scanner,

ZAP tools. They compared the results among these tools and found that ZAP is less efficient [23].Based on clustering techniques R. Akrout et al.,(2014) presented a methodology aiming to identify web application vulnerability. They developed Wasapy vulnerability scanner and compared the result with W3af 1.1, Skipfish 1.9.6b, and Wapiti 2.2.1 focusing on code injection type vulnerability [24]. T. Pandikumar and T. Eshetu.,(2016), presented an approach for automatic penetration testing, based on dynamic analysis for tainted mode vulnerability model problem [25]. F. Lebeau et al.,(2013), proposed an approach from a behavioral model and test patterns, which aims to address both technical and logical vulnerabilities based on Model-Based Vulnerability Testing (MBVT). The research was intended to improve the capability to focus vulnerability testing on the relevant part of the software (e.g., from a risk assessment point of view) and the capability to avoid both false positive and false negative [26]. N. Jovanovic et al.,(2006), addressed problem of vulnerable web application by using static code analysis, context-sensitive data flow analysis, etc. They developed a tool, PIXY for detecting XSS and SQLi vulnerability and also delivered false positive report [27]. J. Fonseca et al.,(2007), proposed a tool that detects two most dangerous web application vulnerability including SQLi and XSS. This tool may identify the web application vulnerability with a large number of interaction into the web browser. They also compared the result by analyzing coverage of vulnerability detection and false positives [28]. H. Jerkovic and B. Sinkovic.,(2017) reviewed on latest web application vulnerabilities into different open source CMS and described the impact of SQLi, XSS and WordPress 4.7 REST-API vulnerability. They also provided a result analysis based on level of risk[29].

In view of the above, it is observed that the research on Word Press CMS was only used for its popularity, flexibility, efficiency, secure development guideline and its third party plug-ins vulnerabilities. Insignificant research focused on Word Press content Injection vulnerabilities. Only H. Jerkovic and B. Sinkovic.,(2017) theoretically discussed about WordPress 4.7 REST-API vulnerability but rigorous analysis conducted on this topic. On the other hand, there were a few other study presented on automated detection tools for web application vulnerabilities like SQLi, XSS, etc. but automated detection of Word Press content Injection vulnerabilities have not been considered. In this paper we will propose a model for automated detection of Word Press content Injection which will be implemented through a tool, named SAISAN.

## 3. OUR PROPOSED MODEL

SAISAN, a detection tool has been implemented on our proposed model for efficient detection of WordPress content injection vulnerability. Exploitation of WordPress content injection vulnerability is performed using http $_POST method. This technique will be applicable for both version 4.7.0 and 4.7.1. The technique allows an attacker to inject the content of the respective vulnerable WP web application page by sending a simple http $_POST request. This proposed model architecture and work-in-procoess are presented below:

Step 1- By using online WP developers modules, our tool will identify the WordPress CMS versions. It will verify whether the versions are 4.7.0 and 4.7.1 or not.

Step 2- Search for Vulnerable Page: As soon as the WordPress version matches with the given versions i.e. 4.7.0 or 4.7.1, SAISAN will send another request for searching the default content page i.e. json/wp/v2/posts/ from the respective web application. Availability of the above page from the given WP version will confirm the WordPress content injection vulnerability. The tool will store the URL link of the page in a variable for providing the output.

Step 3- Output: The tool, SAISAN will show the URL link as output if the vulnerablilty exists in the web application. Otherwise, it will display an error messege of "WordPress Content Injection vulnerability not found".

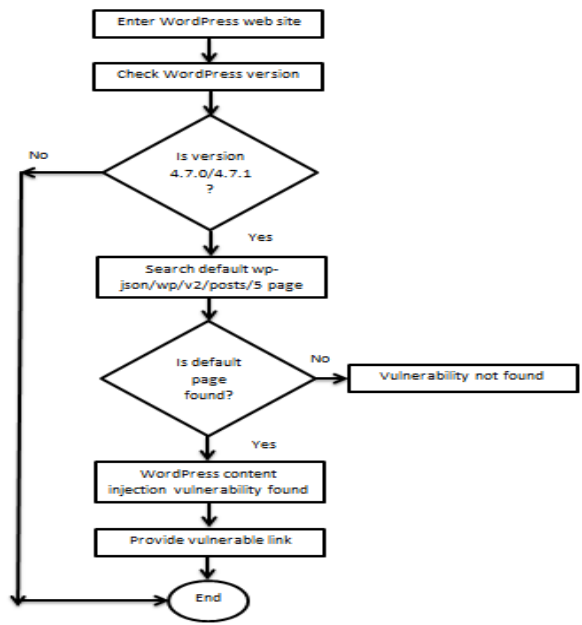

Figure 2. Proposed Detection Model for WP Content Injection Vulnerability

## 4. WP CONTENT INJECTION PROCESS

According to our methodology, we have already discussed that CMS version need to be identified first. Default injectable content page (i.e. /wp-json/wp/v2/posts/) need to be detected to confirm the given vulnerability.

http://anywordpresssite.com/index.php/wp-json/wp/v2/posts/

If the following content is found into the system, it will allow an attacker to inject any content in the respective page. While accessing any content page of the following post directory, it will be shown as below broken html pages:

""""[{"id":123907,"date":"2017-03-29T13:52:56","date_gmt":"2017-03-29T20:52:56","guid":{"rendered":"https:\/\/www.anywordpresssite.com\/?p=123907"},"modified":"2017-03-29T13:53:12","modified_gmt":"2017-03-29T20:53:12","slug":"yubihsm-2-open-beta- """

From the above output, it discloses some ID perameter against some contents or posts. By sending the following http $_POST request to any specific ID perameter along with any content as the attackers like, the vulnerable page will then update with the injected content.

{"id" : "123907Test", "title" : "Checking for injection", "content" : "here is the vulnerable page"}

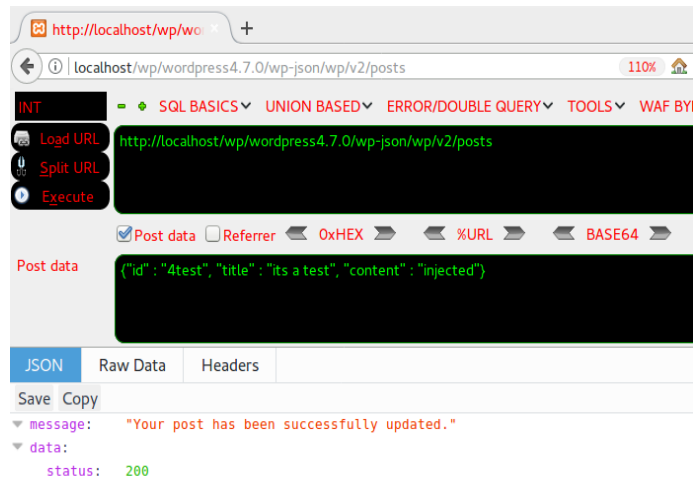

Figure3. Manual Injection Process

If the injection is performed successfully, the injected content should be displayed into the post page. E.g: http://anywordpresssite.com/index.php/wp-json/wp/v2/posts/?ID=123907 Testpage.

## 5. EFFECT OF WP CONTENT INJECTION VULNERABILITY

WordPress content injection vulnerability is Privilege escalation vulnerability which allows an unauthenticated user to modify/inject the content of any post or page within a WordPress site. This privilege escalation vulnerability affects the WordPress REST API that was recently added and enabled by default on WordPress 4.7.0. One of the REST endpoints allows access (via the API) to view, edit, delete and create posts. Within this particular endpoint, a subtle bug allows visitors to edit any post on the site. The REST API is enabled by default on all sites using WordPress 4.7.0 or 4.7.1. If your website is running on any of the above WP versions, it means the page have the bug of content injection.

### 5.1. Effect on REST API

In wordpress CMS version 4.7.0 has a default function i.e register_rest_route() in "/wp-includes/rest-api/endpoints/class-wp-rest-posts-controller.php" directory which generates the ID perameters to set as numeric vaule. Therefore, at the endpoint /wp/v2/posts/ or any type of content can be modifyed, edited and also deleted. If the injecting header content is 12345 for an example, the endpoint of the URL perameter will be like, /wp/v2/posts/12345. The register_rest_route() function is describe below:

However, the numeric value of the id perameter may be rejected by the endpoint_controller in some cases. The REST API always accepts the alphameric values and these are the main reason to exploit content injection vulnerability. For an example,

*index.php/wp/v2/posts/12345?id=12345Testpage*

International Journal on Cybernetics & Informatics (IJCI) Vol. 6, No.3/4/5, October 2017

```
public function register_routes() {
    register_rest_route( $this->namespace, '/' . $this->rest_base, array(
        array(
            'methods'             => WP_REST_Server::READABLE,
            'callback'            => array( $this, 'get_items' ),
            'permission_callback' => array( $this, 'get_items_permissions_check' ),
            'args'                => $this->get_collection_params(),
        ),
        'schema' => array( $this, 'get_public_item_schema' ),
    ) );

    register_rest_route( $this->namespace, '/' . $this->rest_base . '/(?P<type>[\w-]+)', array(
        array(
            'methods'  => WP_REST_Server::READABLE,
            'callback' => array( $this, 'get_item' ),
            'args'     => array(
                'context' => $this->get_context_param( array( 'default' => 'view' ) ),
            ),
        ),
        'schema' => array( $this, 'get_public_item_schema' ),
    ) );
}
```

Figure 4. Rest API

In this phase, user will pass the value to the update_item_permission_check() funtion and this will allows user to modify/inject the posts without any proper authentication as wordpress application always carries this type of id perameter value, and passing the value will modify the post contents using update_item() function.

## 5.2. Effect on Web Application

As we mentioned, WordPress content injection vulnerability is a privilege escalation bug that allows the unauthorized user to edit any post or content of a WP site. The impact and consequences of WordPress content injection attacks can be classified as follows:

1. Confidentiality: Loss of confidentiality is a major problem with WordPress content Injection attacks as an unauthorised person will get the access to the sensitive data with the privileges of editing and altering, which may proceed to disclosure of privacy and confidentiality.

2. Integrity: Keeping data unchanged and intact is a major concern and this may be violated due to a successful WordPress content injection attack, whcih will allow external source to make unauthorized modifications such as altering or even edit information from target web sites. This will destroy the integrity of the valuabe data.

3. Authorization: Successful exploitation of WordPress content injection vulnerability, allows attacker to change authorization information and gain elevated privileges.

## 6. METHODOLOGY

In this section, we have described our key contributions for this paper which are the detection model for WP content injection and a tool (SAISAN) prepared as per the proposed model. Our detection tool was developed to detect content injection vulnerability in WordPress version 4.7.0 and 4.7.1. This tool is developed in Python on Linux platform. The target audiences of this tool are information security specialists, researchers, penetration testers and IT practitioners. Steps of the given tool to detect the above vulnerability are discussed below:

The implementation of SAISAN and code analysis are described in the following subsections:

1. API response collection: Send a request to Online WP Developer's Module with the target web application's URL for identifying the version of WordPress CMS. The module will response with the version to SAISAN with the output of the WordPress version.

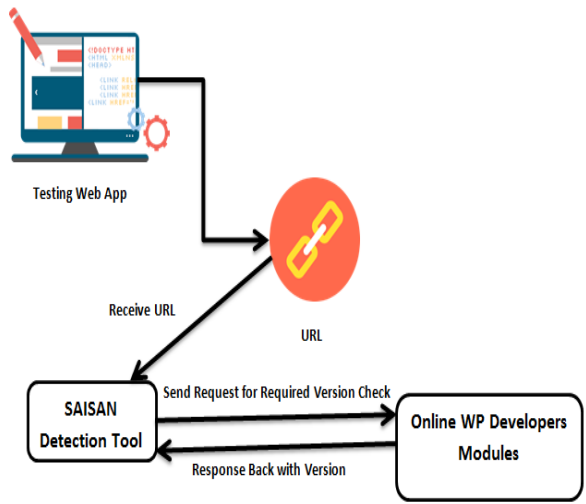

Figure 5. Response collection process

2. Code Analysis: From our previous discussion we know that our tool identifies the suspect vulnerable application based on the CMS version. If it matches with any of the affected versions, it searches for the injectable content page of default post directory. The structural procedure of WP content injection vulnerability detection are given below:

   1. Step 1: (Taking input)
        receiving the URL using input function
            return URL
   2. Step 2: (Checking wordpress version)

        Checking wordpress version using web_api_response
        if version match
            the version is infected
            then searching for the vulnerable_page
        else not infected
            end of condition
   3. Step 3: (Injection testing)
        injectable page finding by matching with the default vulnerable_page
        vulnerable_page = location of content

        if page found
            END
3. If the vulnerable page is directory found, the Tool SAISAN confirms the availability of WP content injection vulnerability and provides the exact vulnerable link. (Figure 6)

Figure 6. Screen shot that shows message of "vulnerability found" with the WP version

If the vulnerability is not found, SAISAN detection tool provides the output box that shows vulnerability is not found. (Figure 7)

Figure 7. Screen shot that shows message of "vulnerability not found" with the WP version

## 7. RESULT ANALYSIS

In this research, we have formulated the sample size mechanism using a universal calculator, provided by G*Power 3.1.9.2. In this study, we examined web applications that developed by WP on April 2017. We have used SAISAN detection tool which follows black box testing method. We have analyzed this dataset based on the following four criterions:
1. Result based analysis;
2. Version wise vulnerability;
3. Sector wise vulnerability ;
4. SAISAN vs. Manual test.

### 7.1 Result Based Analysis

The figure 8 shows that among 176 WP web applications, 59 were found vulnerable. It is observed that almost 34% of WP websites still contain WP content injection vulnerability.

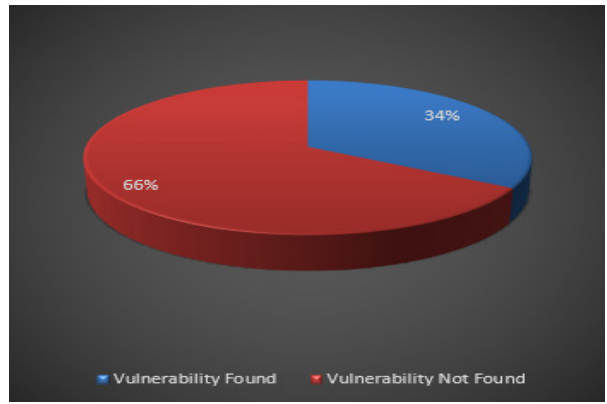

Figure 8. WordPress Content injection vulnerability found and not found

## 7.2 Version Wise Analysis

In our study, content injection vulnerability occurs of WP two vulnerable versions, 4.7.0 and 4.7.1. In this review, we have found 59 CMS web applications are vulnerable where 59% of web applications are built on version 4.7.0 and 41% are built on version 4.7.1.

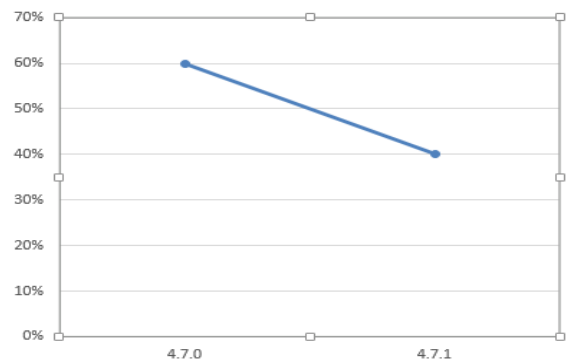

Figure 9. Version based vulnerability graph

## 7.3 Sector Wise Analysis

We randomly select 176 WP web applications where 59 CMS web application contains this injection vulnerability. We have analyzed this data set against five sectors. The sectors are: educational sector, financial sector, medical sector, online portal and blog. Among those five sectors, we have found education sector has the most critical condition. 51% CMS of educational web application have this vulnerability. Among the other sectors, medical sector has 8%, online portal has 15% and blog has 26% vulnerability. We did not find the given vulnerability in financial sector.

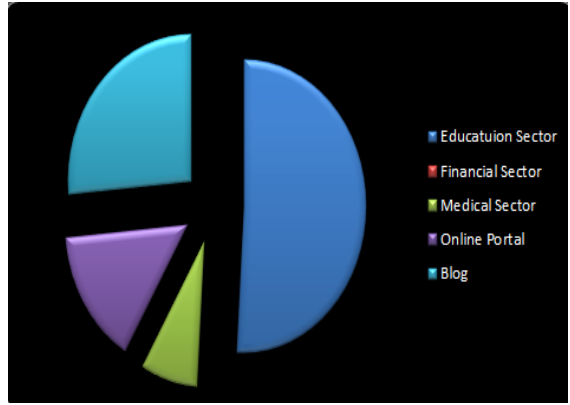

Figure 10. Sector Based Analysis with the percentage of Vulnerability

### 7.4 SAISAN vs Manual Analysis

In the SAISAN we have conducted black box penetration testing to the web applications for getting the result of the existence of content injection vulnerability. However, we have tested the web application as a general user and look for that default file path using manual black box testing. If the vulnerable page existed, we tried to inject manually using Firefox's popular official add-on known as "Hack bar". Hack bar allows users to modify the post based traffic. Based on that request and response we got confirmation of the whole injection process. On the other hand, our developed tool SAISAN identified the CMS version of application first using the online word-press resources. Our tool will send another request for searching the default vulnerable content page i.e. json/wp/v2/posts/ from the respective web application.

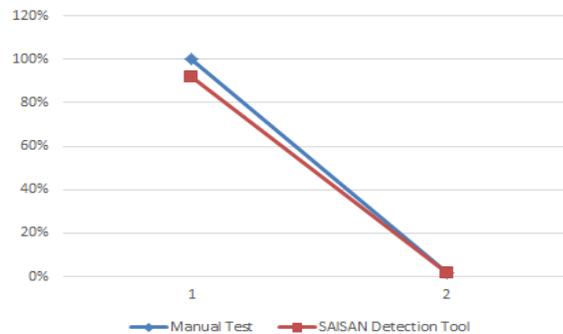

Figure 11. Manual VS SAISAN detection tool

If the given page found, it will store in the URL and display as output. After comparing with the manual black box testing, we found 92% accuracy in out tool, SAISAN.

## 8. DISCUSSION

Based on our developed tool, we have led to an analysis performed on 176 random selected WP CMS web applications. In our survey we found that 34% WP web applications still have WP content injection vulnerability. Those vulnerable web applications are divided into five sectors where we have revealed that education sector is in the most danger position having 51% of the same vulnerability whereas financial sector is holding in a very good position having no WP content injection vulnerability.

To demonstrate the effectiveness of our implemented tool, we did a survey on manual detection using SAISAN where it shows that our tool detected the vulnerability with the accuracy of 92%.

Content injection allows the users to inject malicious code or scripts by updating the pre-existed pages without following a proper authentication to the users. Therefore, an attacker can inject malicious codes to that specific page within very short period of time and getting full access to the web application when the attacker is sending post requests with malicious shell codes for getting control over the web application.

## 9. GENERAL PREVENTION TECHNIQUE

As WordPress is a public CMS, it always recommend administrator to update the version of 4.7.0 and 4.7.1 as it has the content injection vulnerability. If the web administrator uses the latest version, this type of problems can be solved. Most of the cases, third party plugins get different vulnerabilities. Respective administrator should install paid plugins instead of free one.

## 10. CONCLUSIONS

Web application is the basic tool to automate the activities and provide services to humankind. CMS based tools are popular to develop these web applications as they are easy to manage. However, this content injection vulnerability in some popular CMS may create disorder in the daily life activities, which are dependent on such web applications and these applications may become very hard to manage. Therefore detection of the vulnerability and taking necessary measures against the detected vulnerability may save the services these applications provide. In order to detect WP content injection vulnerabilities, this paper proposes a detection model, and developed a tool, SAISAN based on it. After performing an examination on 176 WP managed web application, we found that WP content injection vulnerability is common in two versions of WP i.e. version 4.7.0 and 4.7.1 where the default page directory –json/wp/v2/posts/ existed with privilege escalation bug by which unauthorized user can modify the content of any post or page. In our analysis we compared the result of the above vulnerability detection by using manual black box testing verses our detection tool, SAISAN and get the result of 92% accuracy. Our research is an ongoing work. At this stage, our tool only can detect WP content injection vulnerability. In future we incorporate other vulnerabilities (such as SQLi, XSS, RCE and LFI etc.) detection in our tool.

### ACKNOWLEDGEMENTS

We thank the authorities of the organisations who have given us permission to examine their websites. We also acknowledge the authority of Cyber Security Centre, Daffodil International University for their continuous support and cooperation to continue our research work.

International Journal on Cybernetics & Informatics (IJCI) Vol. 6, No.3/4/5, October 2017

## AUTHORS

**Md. Maruf Hassan** was born in Bangladesh, in 1983. He received his Bachelor Degree in Information Systems and Masters in Computer Science and Engineering. Currently, he is working as Assistant Professor in the Department of Software Engineering and also serving as Assistant Director in Cyber Security Centre at Daffodil International University. Research interests include Information Security, Cyber Security, Malware Detection and Machine Learning.

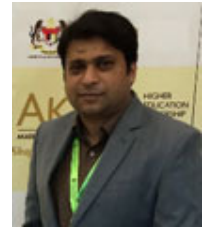

**Kaushik Sarker** was born in Bangladesh, in the year 1988. He received his B.Sc. in Electronics and Telecommunication Engineering under Electrical Engineering and Computer Science and M.Sc. in Computer Systems and Network Engineering under Computing and Information Systems. Currently working as a Senior Lecturer & Associate Head in the Department of Software Engineering, and he is the project Lead of Embedded Gas Engine Generator System at Daffodil International University, Dhaka. His research interests include Digital Image Processing and Computer Vision, Robotics, Embedded Systems, Artificial Intelligence, Data Science and Cyber Security.

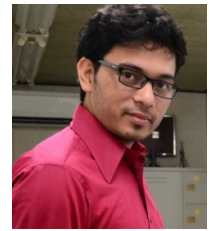

**Saikat Biswas** was born in Bangladesh, in 1994. Now, he is pursuing his under graduate degree from the Department of Software Engineering at Daffodil International University (DIU). Currently he is working as a researcher in Cyber security Centre, DIU. Research interests include Information Security, Cyber Security.

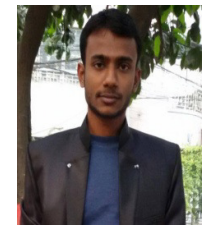

**Md. Hasan Sharif** was born in Bangladesh, in the year of 1994. Now, he is pursuing his under graduate degree from the Department of Software Engineering at Daffodil International University (DIU). Currently he is working as a researcher in Cyber security Centre, DIU. Research interests include Information Security; Cyber Security.

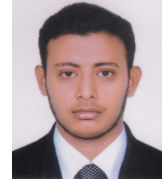